\documentclass{aa}
\newcommand{\mctw}[1]{
\multicolumn{2}{c}{#1}
}

\def\et{et al.}
\def\aj{AJ}

\def\apj{ApJ}

\begin{document}

   \thesaurus{
03            
              (03.20.2;  
               11.10.1;  
               11.17.4: MG\,J0414+0534;  
               12.07.1)} 

   \title{VLBI imaging of the gravitational lens MG\,J0414+0534}


   \author{E. Ros \inst{1}
           \and
           J.C. Guirado \inst{2,3}
           \and
           J.M. Marcaide \inst{2}
           \and
           M.A.\ P\'erez-Torres \inst{2}\thanks{Present address:
             Istituto di Radioastronomia, Via Gobetti 101, 
             I-40129 Bologna, Italy}
           \and
           E.E.\ Falco \inst{4}
           \and
           J.A.\ Mu\~noz \inst{4,5}
           \and
           A.\ Alberdi \inst{6}
           \and
           L.\ Lara \inst{6}
          }

   \offprints{E.\ Ros, ros@mpifr-bonn.mpg.de}

   \institute{Max-Planck Institut f\"ur Radioastronomie,
              Auf dem H\"ugel 69, D-53121 Bonn, Germany
          \and
              Departament d'Astronomia i Astrof\'{\i}sica,
              Universitat de Val\`encia, E-46100 Burjassot (Val\`encia), Spain
          \and
              Centro de Astrobiolog\'{\i}a, Instituto Nacional de
	      T\'ecnica Aeroespacial, Carretera de Ajalvir s/n, E-28850 
              Torrej\'on de Ardoz (Madrid), Spain
          \and
              Harvard-Smithsonian Center for Astrophysics, 
              60 Garden Street, Cambridge, MA 02183, US
          \and
              Instituto de Astrof\'{\i}sica de Canarias,
              V\'{\i}a L\'actea s/n, E-38200 La Laguna (Tenerife), 
              Spain
          \and
              Instituto de Astrof\'{\i}sica de Andaluc\'{\i}a-CSIC,
              Apdo.\ 3004, E-18080 Granada, Spain
             }

    \date{Received 21 January 2000 / Accepted 14 April 2000 }


   \maketitle

   \begin{abstract}

We observed the quadruple gravitationally lensed image of
MG\,J0414+0534 on 23 November 1997
with a global VLBI array at 8.4\,GHz.  We report wide-field imaging 
results of its four components 
at submilliarcsecond
resolution, displaying complex core-like and jet-like extended structures.
A simple model combining a singular isothermal 
ellipsoid to represent the main
lens galaxy, external shear, 
and a singular isothermal sphere to represent an additional, nearby object
accounts well for the core positions and flux densities 
of the VLBI images.
This model predicts delays between the different lensed images of several
weeks.

   \end{abstract}

%

\section{Introduction}

The gravitationally lensed image of QSO 
\object{MG\,J0414+0534} 
($z=2.639\pm0.002$ (Lawrence \et\ \cite{law95}), $V=21.22$,
also PKS\,0411+05, 4C\,05.19, and DA\,131) 
with four components separated by up to 2\,arcseconds
is one of the best examples of morphology produced by
an elliptical gravitational potential
(Hewitt \et\ \cite{hew92},
Katz \et\ \cite{kat97}).
The lens appears to be an elliptical galaxy 
(Schechter \& Moore \cite{sch93}), at a redshift $z=0.9584\pm0.0002$
(Tonry \& Kochanek \cite{ton99}).  

Different studies in radio using the VLA have provided important
information about the structure of the gravitational lens 
(Moore \& Hewitt (\cite{moo97}) and references therein).
Hubble Space Telescope (HST) observations show
an optical arc joining the three brightest components
(Falco \et\ \cite{fal97}).
VLBI images of this radio source
show some extended structure in the sub-images
(Patnaik \& Porcas \cite{pat96}, Porcas \cite{por98}).

We have undertaken a project of multi-epoch observations 
of the quadruple images of each of the gravitationally lensed
objects \object{MG\,J0414+0534} and \object{B1422+231},
to compare possible structural changes in the lensed images 
and measure possible shifts in the relative positions of the components.
These shifts would be caused by intrinsic changes in
the lensed object or by an alteration of the alignment in
the system observer-lens-lensed object due to relative proper motion
of the latter two elements (see Kochanek \et\ \cite{koc96}).
For non-magnified extragalactic objects and
peculiar velocities of, say, 100 to 1000\,km\,s$^{-1}$, 
such relative proper motions would not
be detectable astrometrically over decades. Because of their high
degree of symmetry, quadruple gravitationally lensed systems
can magnify intrinsic velocities 
by factors of 10--100 while minimizing systematic effects for 
measurements of relative motions (Kochanek \et\ \cite{koc96}).

We present here the mapping and modeling results from the first epoch 
of observations
of \object{MG\,J0414+0534}.
We describe our VLBI observations in Sect.~\ref{sec:obs},
report our imaging results of \object{MG\,J0414+0534} in
Sect.~\ref{sec:imag}, discuss our modeling results in Sect.~\ref{sec:model}
and give a summary in Sect.~\ref{sec:summ}.

%

\section{Observations and data reduction\label{sec:obs}}

We observed \object{MG\,J0414+0534} at 8.4\,GHz on 23 November 1997 
with a global VLBI array for over 12\,hr. 
We used a basic integration
time of 0.4\,s.
The antennas used
were (name, 
diameter, location): 
Medicina (32\,m, 
Italy), 
Noto (32\,m, 
Italy), 
Effelsberg (100\,m, Germany), 
DSS63 (70\,m, 
Spain), 
the Very Long Baseline Array (VLBA) 
(10 antennas of 25\,m across the US),
the phased-VLA (130\,m-equivalent, NM, US), 
and
DSS14 (70\,m, CA, US).
Data were 
recorded with right 
circular polarization using mode 128-2-2, achieving a bandwidth of 32\,MHz, and correlated
at the National Radio Astronomy Observatory (NRAO)
VLBA Correlator (Socorro, NM, US).
The observing mode put the correlator close to most demanding performance,
taking advantage of its latest technical improvements (actually, achieving
the smallest sample interval --widest channel bandwidth-- with the
highest spectral resolution --largest delay range-- in the used mode).
The antennas pointed to the center of the field of the
four images of the
radio source (all within the individual antenna beam at the same time) 
and only one correlation 
pass was needed to determine the interferometric visibilities jointly 
for the four components.  Due to operational
reasons, the strongest of the four components (A1, see below)
was chosen as phase center at correlation time.

The observing duty cycle consisted of 2\,min on the
calibrator source \object{TXS\,0357+057}
and 5.5\,min on the target source \object{MG\,J0414+0534}.
We allocated some calibrators and fringe-finders (\object{BL\,1803+784}, 
\object{QSO\,1928+738}, \object{BL\,2007+777}, 
\object{3C\,84} and \object{3C\,345}) at 
the beginning and the end of the observations.  

We analyzed the data using
the Astronomical Image Processing System
({\sc aips}, 
version {\sc 15oct99}).
We applied a priori amplitude calibrations determined from the
measured system temperatures and available antenna gains.  
We aligned the phase slopes
across each of the 16\,MHz bands using the 
information from 
each telescope, when available. In the absence of this
information, we applied manual phase
calibration deduced from observations of \object{TXS\,0357+057}, 
\object{3C\,84}, and \object{3C\,345}.
We then performed standard fringe-fitting on the calibrators and
the target source.  
We made
a preliminary map
of \object{MG\,J0414+0534}
by down-weighting  (tapering) the longest baselines
to gain sensitivity to extended surface brightness.
We used this initial image to perform a second fringe-fit data correction
in {\sc aips}.

Since the total angular size of \object{MG\,J0414+0534}
is 2\,arcseconds, 
wide-field imaging techniques (see, e.g., Garrett \et\ \cite{gar99}) 
are needed.
At a given frequency, the field of view of a VLBI map 
is determined
by the averaging time of the data and the bandwidth of the observation
(see, e.g., Cotton \cite{cot99},
Bridle \& Schwab \cite{bri99}). 
Over-averaging in
frequency and time leads to bandwidth- and 
time-smearing, respectively.  
For our data, we 
averaged every four 0.25\,MHz channels at the two
16\,MHz frequency bands to obtain 32 frequency channels, each of
1\,MHz width.
Later on, to speed up
the mapping process, we time-averaged in 4\,s intervals
(shorter averaging times, 
1.2 and 2.4\,s, did not produce significant variations).
The adopted 4\,s and 1\,MHz 
averagings avoid both types of smearing, allowing a 
size of the data set just small enough to be manageable with our 
computing resources. This wide-field mapping 
technique allows to image fields 
of arcsecond-size with milliarcsecond resolution. 
However, a single
mapping field results of unmanageable size.
As a solution, the {\sc aips} task {\sc imagr}
provides a reliable wide-field mapping methodology based on a
multi-windowing approach.  
Our imaging strategy emphasized tapering
at the beginning of the mapping process to help recovering
most of the extended flux density in the sub-images.  
In further iterations
we lifted the tapering to achieve the full
resolution provided by the global VLBI array.  
Using a hybrid weighting (between uniform and natural, {\sc robust=$-$1} in
{\sc imagr} task) at full resolution, the synthesized beam obtained 
was of 2.55$\times$1.13\,mas (position angle (P.A.) 14.6$^\circ$).  

\section{Imaging results\label{sec:imag}}

\begin{figure*}[phtb]
\vspace{510pt}
\includegraphics{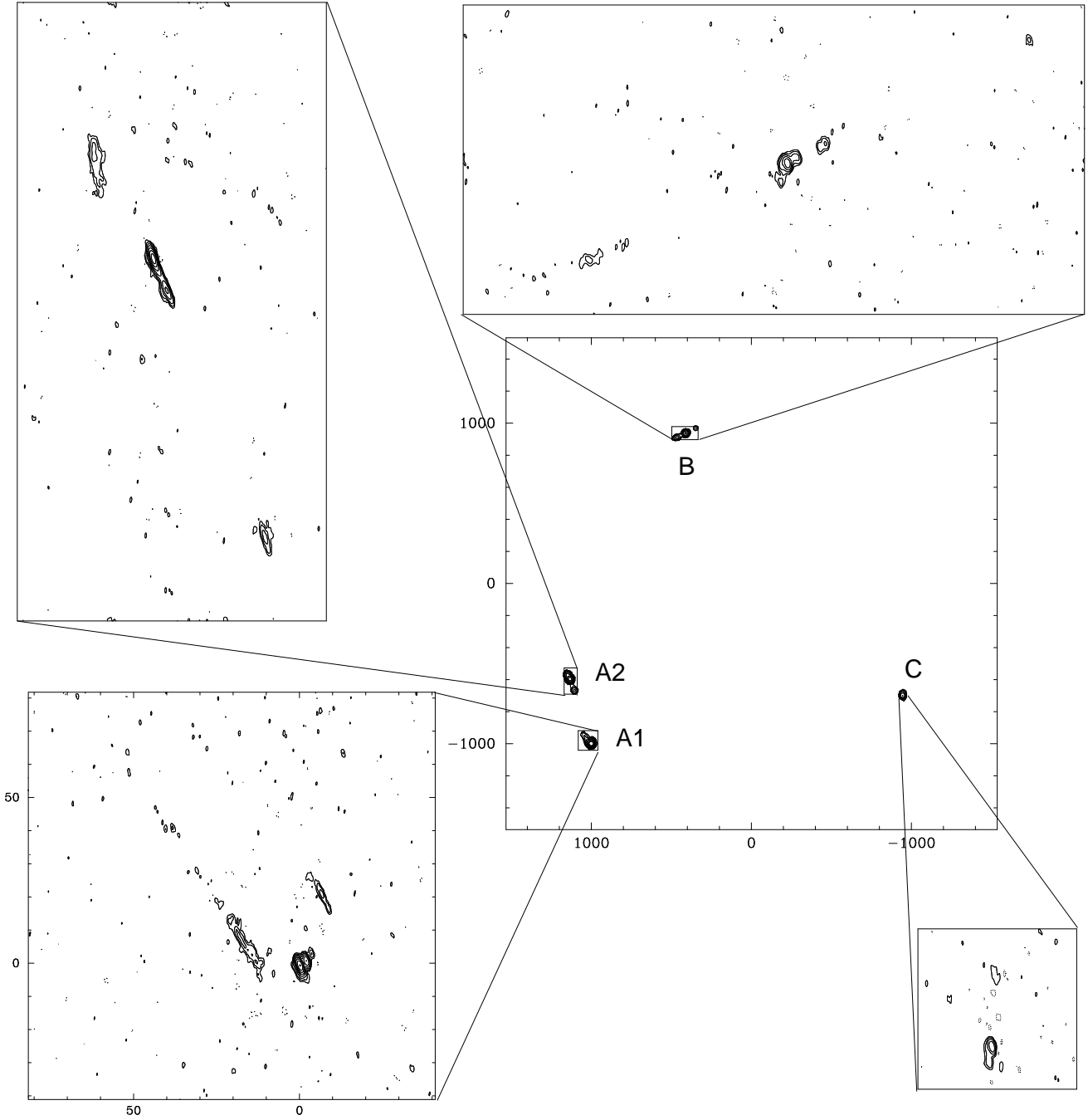}
\caption{Composition of 
our 
images of the four 
gravitationally lensed images, labelled as A1 (bottom, left), A2 (top,
left), B (top, right), and C (bottom, right) 
of \object{MG\,J0414+0534} 
at 8.4\,GHz with
a global VLBI wide-field image (central panel) of the radio source.
The axes are relative right ascension and declination, in milliarcseconds. 
The wide-field image represents a combination of the four mapped 
windows (or sub-images) in a wider grid convolving their
{\sc clean}-components with a circular beam of 25\,mas 
in diameter.  It displays contours
of 1\,mJy/beam$\times$$(-1,1,2,4,\cdots)$.  The 
sub-images are convolved with their synthesized beam of 2.55$\times$1.13\,mas
(P.A.\ 14.6\degr).
The contours are
0.5\,mJy/beam$\times$$(-1,1,2,\cdots)$.  The root-mean-square (rms) noise
levels in the sub-images were 0.25, 0.21, 0.16, and 0.16\,mJy/beam,
for A1, A2, B, and C, respectively.  The values of the
peaks of brightness of each of the sub-images are
given in Table~\ref{table:comp-summ}.  
}
\label{fig:maps}
\end{figure*}

\begin{figure*}[htbp]
\vspace{302pt}
\includegraphics{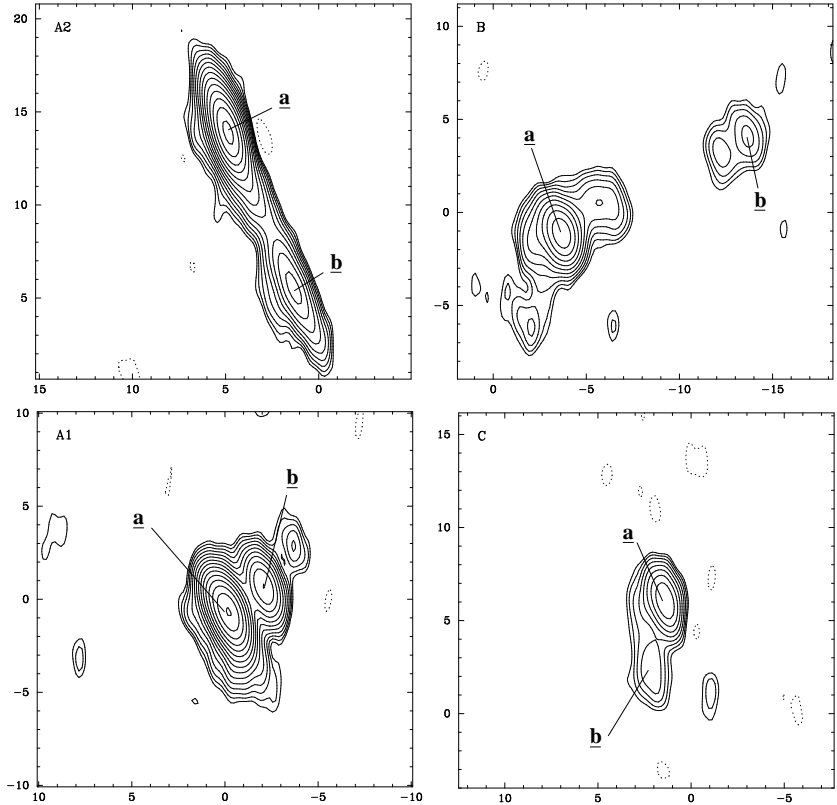}
\vspace{-30mm}
\hfill      \parbox[b]{5.5cm}{\caption[]{
Sub-image cores
convolved each with the appropriate beam $\sim$2.55$\times$1.13\,mas, 
P.A.\,$14.6\degr$.
The axes are relative right ascension and declination, in milliarcseconds. 
Contours are 0.6\,mJy/beam$\times(-1,1,\sqrt{2},2,\cdots)$.
{\bf \underline{a}} and {\bf \underline{b}} in the
maps indicate the
two brightest features in each sub-image .
\label{fig:cores}
}}
\end{figure*}

We show the final {\sc aips} maps in Fig.~\ref{fig:maps}.  
At the center of Fig.~\ref{fig:maps}, we plot a wide-field image 
of the source, restored
with a circular Gaussian beam 25\,mas in diameter 
to illustrate
the location of the four components.
At this resolution,
\object{MG\,J0414+0534} 
comprises two bright, close components (A1 and A2) separated by 
0$\rlap{.}^{\prime\prime}$47
and two weaker components B and C
separated 2$\rlap{.}^{\prime\prime}$12 from each other, 
and 2$\rlap{.}^{\prime\prime}$02 and 2$\rlap{.}^{\prime\prime}$08 
from A1, respectively.
We show the map dimensions in the panel corresponding
to A1 (bottom, left), preserving the same map scale in the other three 
sub-images.
Each of these sub-images has been restored with the appropriate
synthesized beam, which is about $2.55\times1.13$\,mas, P.A.\ $14.6\degr$.
In Fig.~\ref{fig:cores}, 
we show more detailed maps of the
brightest features of the four sub-images at full resolution.
We present in Table~\ref{table:comp-summ}
their peaks of brightness $I_{\rm peak}$
and the total flux density values ($S_{\rm tot}$) for the whole sub-image
and for core-like and jet-like structures, obtained computing the
flux densities of the {\sc clean} components in each of those regions.
VLA flux density
ratios among sub-images given by Katz \& Hewitt (\cite{kat93}) and Katz \et\ 
(\cite{kat97}) are similar to ours.
Our results show A1 and A2 somewhat brighter, and B somewhat
weaker as their results.  It is difficult to compare both 
data sets, since some flux density
present at the VLA may not be detected in the VLBI images.

%

%

%
\begin{table}[htbp]
\caption{Total flux density values for each of the sub-images and their 
sub-components in Fig.~\ref{fig:maps}. 
\label{table:comp-summ}}
\begin{tabular}{@{}l@{\,\,}r@{\,\,}r@{\,\,}c@{\,\,}c@{\,\,}c@{\,\,}r@{:}l@{\,\,}r@{:}l@{\,\,}r@{$\pm$}l@{\,\,}@{}}
   &           &          &      &       & {{\sc core}$^{\rm c}$} 
                                                 & \mctw{{\sc jet}$^{\rm d}$} 
                                                             & \mctw{{\sc cjet}$^{\rm d}$} \\
Id & $x$$^{\rm (a)}$ 
               & $y$$^{\rm (a)}$ 
                          & $I_{\rm peak}$ 
                                 & $S_{\rm tot}$$^{\rm (b)}$
                                         & $S_{\rm tot}$
                                                 & \mctw{$S_{\rm tot}$} 
                                                             & \mctw{$S_{\rm tot}$} \\
   & {\tiny [mas]} 
               & {\tiny [mas]} 
                          & {\tiny [mJy/beam]} 
                                 & {\tiny [mJy]} 
                                         & {\tiny [mJy]}     
                                                 & \mctw{\tiny [mJy]}   
                                                             & \mctw{\tiny [mJy]}      \\ \hline
A1 &     $0.00$ &     $0.00$ & 65.5 & 155.6 & 115.6 & {\footnotesize NE} & 33.6 & {\footnotesize N}  &  7.4 \\ 
A2 &  $+133.96$ &  $+405.40$ & 50.2 & 129.3 &  97.0 & {\footnotesize  N} & 16.5 & {\footnotesize S}  & 10.2 \\ 
B  &  $-588.52$ & $+1937.52$ & 14.9 &  50.0 &  34.0 & {\footnotesize SE} & 10.4 & {\footnotesize NW} &  1.0 \\ 
C  & $-1945.40$ &  $+299.72$ &  7.9 &  21.6 &  11.8 & {\footnotesize  N} &  2.1 & \mctw{---} \\ \hline 
\end{tabular}
\begin{footnotesize}
\begin{list}{}{
}
\item[$^{\rm a}$] Relative position of the peaks of brightness on the
maps. The image pixel size is 0.16\,mas. 
\item[$^{\rm b}$] Total flux densities in the fields shown
in Fig.\ \ref{fig:maps}.
\item[$^{\rm c}$] Total flux densities for the complex core-like region
shown in Fig.\ \ref{fig:cores}.
\item[$^{\rm d}$] Total flux densities for the jet-like structures,
labelled as {\sc jet} and {\sc c(ounter)-jet}.
\end{list}
\end{footnotesize}
\end{table}

The gravitational lensing geometry produces four images of the same object, 
deformed and with different magnification factors.  
In addition, some lensed, unresolved components
might also be fainter than the sensitivity level of our maps,
causing some confusion in the feature identification among
different sub-images. 
In our maps, we do not observe isolated, compact components
(like in the quadruple lensed object \object{B1422+231}) but, rather,
very complex structures, with core-like and jet-like regions that show
different morphologies for each lensed image.  

We discuss in turn the features of each lensed image.
{\it Image A1.}
It is the most complex
and the brightest.
There are two parallel jet-like components, totally misaligned
with respect to the core-like region.  
The fainter one is 20\,mas away, P.A.\ 
$\sim-15^\circ$.  The brighter one, very stretched along P.A.\ $\sim30^\circ$
begins at about 10\,mas southwest of
the core-like region, and extends over more than 50\,mas.
The brightest feature (core-like) has two components along
P.A.\ $\sim-60^\circ$ and a resolved structure.
{\it Image A2.} 
Two extended regions not quite co-aligned 
(in the same direction) are present at about
30\,mas to the NE and 80\,mas to the SW of the central (core-like) area. 
The core-like area 
(Fig.~\ref{fig:cores}, top, left) is symmetric, with two
components, elongated over more than
15\,mas at P.A.\ $\sim20^\circ$.
{\it Image B.}
It presents two extended jet-like regions, fairly aligned with its 
central area, at about 70\,mas to the NW and SE.
The core-like area itself
is double and elongated more than 15\,mas at P.A.\ $\sim-70^\circ$.
{\it Image C.}  It presents a double N-S structure,
the northern feature being the brightest, and extended emission to the
north of this double structure.
The core-like region presents 
some extended emission to the South.

A comparison of our maps with previously published images of 
\object{MG\,J0414+0534} shows that the EVN 
images at 5\,GHz shown in Porcas (\cite{por98})
are similar to our maps, but with a wider beam and lower dynamic
range.  The same holds for the 1.7\,GHz EVN images from Patnaik \& Porcas
(\cite{pat96}).
The optical image from
Falco \et\ (\cite{fal97}) shows emission
extending over an arc joining A1, A2 and B.

\section{Lens models\label{sec:model}}

Quadruple gravitational lens systems can be produced by an elliptical
lensing potential.  Many attempts have been made to model
\object{MG\,J0414+0534} (e.g.\ Kochanek \cite{koc91}, Hewitt \et\
\cite{hew92}, Witt \et\ \cite{wit95}).  Falco \et\ (\cite{fal97})
could not find models that were simultaneously consistent with the
brightest features of the optical images
and the lower-resolution radio data from
the VLA (e.g., Katz \& Hewitt \cite{kat93}).
The image flux densities at radio wavelengths have shown little
variation, and have not yielded time delays (Moore \& Hewitt
\cite{moo97}).
We show below that the
addition to the lens model of a previously known but unsuspected lens
galaxy near the images suffices to account for the properties of the
images.  This new model also predicts
time delays between the different lensed sub-images.

\begin{figure}
\vspace{231pt}
\includegraphics{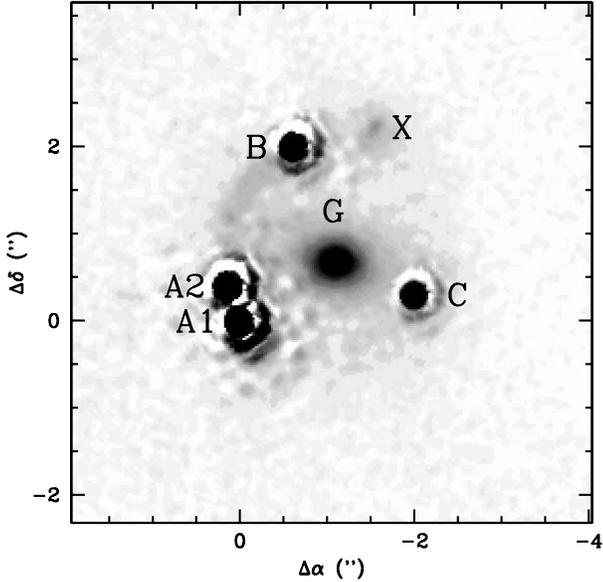}
\caption{CASTLES HST/NICMOS image of \object{MG\,J0414+0534} at H-band ($\lambda=1.6\,\mu$m),
after deconvolution of the point-spread function.
\label{fig:hst}}
\end{figure}

We fitted a simple lens model to our VLBI data.  Because the lens
galaxy G is not detected in these data, we used the deep HST NICMOS
H-band ($\lambda=1.6\,\mu$m) image from the CfA/Arizona Space
Telescope Lens Survey (CASTLES; e.g., Falco \et\ \cite{fal99}) to
estimate the coordinates of the center of brightness and the
structural parameters of the lens galaxy G (see Fig.~\ref{fig:hst}).
We registered the VLBI and point-like infrared images by assuming that
the infrared positions of A1, A2, B and C correspond to the
peaks of brightness of each of the four VLBI sub-images.
The transformation that maps the infrared
coordinates to those of {\bf \underline{a}} or {\bf \underline{b}} is
a scale change (to take into account a chip position and
wavelength-dependent uncertainty of $\sim 10$\,mas in the NICMOS/NIC2
scale) with a rotation (to take into account the uncertainty of $\sim
0\fdg1$ in the orientation of HST). We found the transformation to
the feature {\bf \underline{a}} ({\bf \underline{b}}) requires a rotation of
$0\fdg28$ ($0\fdg16$) and a scaling by $0.992$ ($0.991$), with rms
residuals of 11~(5)\,mas. The radio and infrared emission do not
emanate necessarily from the same region of the source.  Therefore, we
adopt the brightest feature {\bf \underline{a}} as the 8.4\,GHz
counterpart of the quasar, and assign to it positional uncertainties
of 11\,mas, to reflect the uncertainty in the registration.
 In this paper, we ignore the lensed jet structure because of
the difficulty in establishing the correspondence between lensed
components along the jets.  After ongoing software developments are
completed, this aspect will be addressed in a forthcoming paper.

Taking as origin of coordinates the feature {\bf \underline{a}} on image
A1, we use as constraints for the lens models the 6 relative
coordinates of feature {\bf \underline{a}} (Table~\ref{table:comp-summ})
(assuming a conservative 0.5\,mas uncertainty in the positions),
the coordinates of G, $x_G = -1072\pm11$\,mas, $y_G = +665\pm11$\,mas,
and the 3 peak of brightness ratios from the radio sub-images
(with 10\% uncertainties due to the possibility of microlensing).
We adjusted lens models with a small number of
parameters.  Our initial
model for the mass density distribution is a singular isothermal
ellipsoid (SIE), with 5 parameters: the Einstein ring radius $b$, the
ellipticity $\epsilon$, sky position angle $\theta_\epsilon$, and
coordinates $x$ and $y$.  SIE models yield values that are consistent
with the masses of other lens galaxies (e.g., Leh\'ar \et\
\cite{leh00}) and that reflect the properties of elliptical galaxies
(e.g., Kochanek \et\ \cite{koc00}).  As a rule, single mass
distributions require external lensing shear to account for the
properties of quadruple lens systems (Keeton \et\
\cite{kee97}). Therefore, we added external shear with amplitude
$\gamma$ and position angle $\theta_\gamma$ to our model. The total
number of parameters of our SIE+$\gamma$ model is 7.

We adjusted all 7 parameters of the SIE+$\gamma$ model using the sum
of weighted rms residuals for the peak of brightness ratios, 
the coordinates of G
and those of the {\bf \underline{a}} feature images as the figure of
merit, $\chi^2$. The best-fitting parameters of this initial model
yielded $\chi^2\sim 30$ principally
because the lens fell $\sim 60$ mas away from the observed galaxy.
Therefore, we included object X (Schechter \& Moore \cite{sch93}) at
$x_X = -1457\pm13$\,mas, $y_X = +2122\pm13$\,mas (derived from the
CASTLES data) as a possible additional component of the lens.  Because
X is about 10 times fainter than G, its structure cannot be
distinguished from a circular brightness distribution.  Thus, we
simply assumed X is a singular isothermal sphere (SIS) and added its 3
parameters, $x_X$, $y_X$ and $b_X$ to the model,
where $x_X$ and $y_X$ are constrained by the CASTLES data.
 We adjusted the
resulting 10 parameters of the SIE+$\gamma$+SIS model, which therefore
has 3 degrees of freedom (DOF).  We thus found values that account nearly
exactly for all the positions of the images of feature {\bf
\underline{a}} as well as those of G and X, 
and also for the ratios of the peaks of brightness at the
core-like regions, with a final
merit per DOF $\chi^2/3=1.7$ (arising solely from the constraints
in the peak of brightness ratios).
The parameters of this model are shown in Table~\ref{table:model}.
The model predicts that B is the leading image when the source varies:
the time delays in a flat cosmology ($\Omega_0=1$,
$H_0=65$\,km\,s$^{-1}$\,Mpc$^{-1}$) are
$\Delta t_{A1B}=15.7 \pm 1.3$ days, $\Delta t_{A2B} =16.0 \pm
1.4$ days and $\Delta t_{CB}=66 \pm 5$ days, where the error bars are
an estimation from the variation of the predicted time delay as the
lens model parameters are varied within the $\Delta\chi^2\leq1$ region.

\begin{table*}[htbp]
\caption{Lens model.
\label{table:model}}
\begin{tabular}{@{}cr@{$\pm$}lr@{$\pm$}lr@{$\pm$}lr@{$\pm$}lr@{$\pm$}lr@{$\pm$}lr@{$\pm$}l@{}}
Id & \mctw{$x$$^{\rm (a)}$}
 & \mctw{$y$$^{\rm (a)}$}
 & \mctw{$b$}
 & \mctw{$\epsilon$}
 & \mctw{$\theta_\epsilon$$^{\rm (b)}$}
 & \mctw{$\gamma$}
 & \mctw{$\theta_\gamma$$^{\rm (b)}$} \\
 & \mctw{{\footnotesize [mas]}}
 & \mctw{{\footnotesize [mas]}}
 & \mctw{{\footnotesize [mas]}}
 & \mctw{{(1--$b/a$)}}
 & \mctw{{\footnotesize [\degr]}}
 & \mctw{{(shear)}}
 & \mctw{{\footnotesize [\degr]}} \\ \hline
G & $-1071$&10  &  $+663$&10  & 1090&10  & 0.21&0.05 & --83&6 & 0.096&0.012 &
56&4 \\
X & $-1457$&13  & $+2122$&13  &  180&20  & \mctw{} & \mctw{} & \mctw{} & \mctw{} \\ \hline
\end{tabular}
\begin{footnotesize}
\begin{list}{}{}
\item[$^{\rm a}$] Positional offsets are relative to the peak of brightness in
the A1 image.
\item[$^{\rm b}$] Angles from North through East.
\item[Note:]
Parameter uncertainties were estimated from the range
over which $\Delta\chi^2\leq 1$ as each parameter was varied
and the remaining parameters were optimized.
\end{list}
\end{footnotesize}
\end{table*}

Our simple lens model reproduces successfully the data. It is
remarkable that: a) the position angle of the SIE model is consistent
with that of the light of G ($-96\degr\pm4$); b) the ratio of the 
two Einstein ring
radii, which scales as the square root of the ratio of luminosities,
$b_X/b_G=0.17\pm0.02$, 
is consistent with the brightness ratio of X
and G, whose magnitudes differ by $2.6 \pm 0.5$\,mag; and c) the
amplitude of the shear is also consistent with that expected from
cosmic shear (Bar-kana \cite{bar96}), and with the number of galaxies
seen within a radius of $\sim 1$ arcmin of the lens system (McLeod \et\
\cite{mcl00}).

\section{Summary\label{sec:summ}}

We present detailed state-of-the-art global VLBI images 
of the gravitational lens system
\object{MG\,J0414+0534}.
The individual
images, separated by distances up to 2 arcseconds, exhibit radio
structures extending up to 100\,mas.
The use of wide-field mapping techniques in very
sensitive observations
provides high-quality images of extended structures in
\object{MG\,J0414+0534}, showing features that extend
more than 1500 beam widths across in the E-W direction, 
and more than 750 N-S.

We reproduce successfully the relative positions and
peak of brightness ratios of
the radio cores with a lens model consisting
of a singular isothermal ellipsoid for the galaxy G, an external shear
to it, and a secondary potential given by a singular
isothermal sphere associated with object X.
The model predicts that B is the leading image and C lags all
the other images with delays for all image pairs of several weeks.  
It is important to notice the alignment
of the optical images of the lens galaxy G and the model's 
elliptical potential (SIE).  The model's shear is also consistent in 
amplitude with the expected cosmic shear and 
from the observed environment (within 
a radius of $\sim 1$ arcmin) of the lens system
(McLeod \et\ \cite{mcl00}). 
Our model also provides consistency between the luminosity ratio for X and
G and the ratio of their Einstein radii.
We plan to refine and study further our lens model by adding the
constraints provided by the lensed jets, using a new software package
for lens modeling developed by Keeton \et\ (in preparation) that
avoids the problem of matching components along these jets, in the
different lensed images. 

\begin{acknowledgements}
We are especially grateful to J.\ Romney for essential help provided during
correlation and post-correlation.
We acknowledge also R.W.\ Porcas and A.R.\ Patnaik 
for useful discussions.  This work has
been partially supported by the Spanish DGICYT Grants No.\ PB96-0782
and PB97-1164
and by European Comission's TMR-LSF programme, contract No.\ ERBFMGECT950012.
NRAO is operated
under license by Associated Universities, Inc., under cooperative
agreement with NSF.
\end{acknowledgements}


\end{document}